\documentclass[A4,11pt,english]{article}

\usepackage[T1]{fontenc}
\usepackage[utf8]{inputenc}
\usepackage{geometry}
\usepackage{color}
\usepackage{babel}
\usepackage{float}
\usepackage{amsmath}
\usepackage{amsthm}
\usepackage{amssymb}
\usepackage[noend]{algorithmic}
\usepackage[boxed]{algorithm}
\usepackage[unicode=true,pdfusetitle,
 bookmarks=true,bookmarksnumbered=false,bookmarksopen=false,
 breaklinks=false,pdfborder={0 0 1},colorlinks=false]
 {hyperref}
\usepackage{breakurl}
\usepackage{xspace}

\usepackage{rotating}

\makeatletter

\usepackage{booktabs} 
\usepackage{color} 

\newcommand{\dfr}{\ensuremath{d_{\textup{F}}}}
\newcommand{\Fr}{Fr\'echet }

\begin{document}
\title{A fast implementation of near neighbors queries \\ for Fr\'echet distance (GIS Cup)}

\author{Julian Baldus\footnote{Saarland University,
Saarland Informatics Campus, Saarbr\"ucken, Germany,
\texttt{julianbaldus@gmail.com}}
\and 
Karl Bringmann\footnote{Max Planck Institute for Informatics,
Saarland Informatics Campus, Saarbr\"ucken, Germany,
\texttt{kbringma@mpi-inf.mpg.de}}}
\date{}

\maketitle

\begin{abstract}
This paper describes an implementation of fast near-neighbours queries (also known as range searching)  with respect to the Fr\'echet distance. The algorithm is designed to be efficient on practical data such as GPS trajectories.
Our approach is to use a quadtree data structure to enumerate all curves in the database that have similar start and endpoints as the query curve. On these curves we run positive and negative filters to narrow the set of potential results. 
Only for those trajectories where these heuristics fail, we compute the Fr\'echet distance exactly, by running a novel recursive variant of the classic free-space diagram algorithm.

Our implementation won the ACM SIGSPATIAL GIS Cup 2017\footnote{\url{http://sigspatial2017.sigspatial.org/giscup2017/}}.
\end{abstract}

\newpage

\section{Introduction}

The \Fr distance is a very popular measure of similarity of two given curves $\pi,\sigma$, which roughly speaking measures the minimal length of a leash connecting a dog to its owner as they walk without backtracking along $\pi$ and $\sigma$, respectively. As a natural measure of curve similarity, it has broad applications in geographic information systems.
Alt and Godau introduced the \Fr distance to computer science in seminal papers in the early 90's, where they presented an $O(nm \log (nm))$ algorithm to compute the \Fr distance of polygonal curves $\pi$ and $\sigma$ with $n$ and $m$ vertices~\cite{AltG95,Godau91}. 
Over the decades, \Fr distance developed into a rich field of research, in which many generalizations and variants are studied (see, e.g.,~\cite{AltB10,BuchinBW09,ChambersETAL10,DriemelHPW12}).
However, the quadratic worst-case complexity of Alt and Godau's algorithm is still state-of-the-art, apart from log-factor improvements~\cite{BuchinBMM14}.
The second author recently presented strong evidence that the \Fr distance has no strongly subquadratic algorithms, by proving that any such algorithm would yield a breakthrough for the Satisfiability problem (specifically it would break the Strong Exponential Time Hypothesis)~\cite{Bringmann14}. 

In this paper we study the problem of range searching with respect to the \Fr distance, which comes up in many applications. Here, in the preprocessing phase we are given a set~$D$ of curves, which we call the \emph{database}. As a query we are then given a \emph{query curve} $\pi$ and a \emph{distance threshold} $\delta$, and the task is to output all curves in $D$ that have \Fr distance at most $\delta$ to $\pi$. 

The naive solution is to compute the \Fr distance between $\pi$ and $\sigma$ for each $\sigma \in D$. If $\pi$ has $n$ vertices and $D$ consists of $k$ curves having $m$ vertices each, then this takes time $O(k n m  \log (n m))$. We claim (without proof) that the construction of~\cite{Bringmann14} can be adapted to show that any significant improvement over this running time would yield a breakthrough for the Satisfiability problem, specifically for any $\varepsilon > 0$ an $O((knm)^{1-\varepsilon})$-time algorithm would falsify the Strong Exponential Time Hypothesis. 
However, this only excludes better running time \emph{guarantees in the worse case}, but it does not rule out algorithms that are efficient on realistic instances.

\paragraph{What are realistic curves?}
In the area of computational geometry, researchers tried to formalize what it means for a curve to be \emph{realistic}. Several notions of curves have been developed, such as backbone curves~\cite{AronovHPKW06}, $\kappa$-bounded and $\kappa$-straight curves~\cite{AltKW04}, $\phi$-low density curves~\cite{DriemelHPW12}, and $c$-packed curves~\cite{DriemelHPW12}.  
While these notions improved our understanding of characteristics of curves that make \Fr distance computation hard, they are still very far from modeling real-world trajectories such as the example data set given in the ACM SIGSPATIAL GIS Cup 2017, which is supposedly derived from GPS traffic data in the San Francisco Bay Area. These curves are (close to) shortest paths in an underlying road network, and thus have the following features.

\begin{enumerate}
\item There are few sharp turns (narrow angles) in the input data, or at least every input curve has small \Fr distance to some curve with few sharp turns.
\item Two curves with similar start and endpoints have small \Fr distance.
\end{enumerate}

Note that this is neither a complete set of characteristics of realistic curves, nor a formal definition of a model of such curves. However, it does show some features that are not incorporated in the standard models of realistic curves, or only to a small extend. 

\paragraph{Our Contribution}
In this paper we describe an implementation of \Fr distance range searching, with the goal of having a practically fast method. Our implementation is the winning submission to the ACM SIGSPATIAL GIS Cup 2017\footnote{\url{http://people.mpi-inf.mpg.de/~kbringma/frechetimpl/}}. We also report on preliminary experiments.

Our algorithm consists of three phases. First, we determine all curves in the database whose start and endpoints are sufficiently close to the query curve's start and endpoints. Feature (2) ensures that most of these curves have small \Fr distance to the query curve, and thus the number of listed curves is not much more than the output size. To efficiently enumerate these curves we use a multidimensional analogue of a quadtree. 
In the second phase, we use a greedy algorithm to filter out some of the curves that have sufficiently small \Fr distance. We also have a negative filter that determines some of the curves with too large Fr\'echet distance. Only if both heuristics fail, we have to run an exact Fr\'echet distance decision algorithm, for which we use an novel recursive version of the standard free-space diagram algorithm.

For our GIS Cup submission we also added a simple parallelization over multiple queries, but in this paper we focus on answering a single query.

The first phase (quadtrees) is described in Section~\ref{sec:quadtree}, the second phase (filters) follows in Section~\ref{sec:filters}, and the third phase (exact algorithm) is discussed in Section~\ref{sec:exact}. We conclude with preliminary experiments in Section~\ref{sec:results}.

\section{Preliminaries}

All curves considered in this paper are \emph{polygonal}. 
We define a polygonal curve $\pi$ by its vertices $(\pi_1,\ldots,\pi_n)$ with $\pi_i \in \mathbb{R}^2$ for all $i \in \{1,\ldots,n\}$. We denote by $|\pi| = n$ be the number of vertices of~$\pi$. Moreover, we denote by $\|\pi\|$ the total length $\sum_{i=1}^{n-1} \|\pi_i - \pi_{i+1}\|$, where $\|.\|$ denotes the Euclidean norm. 
We write $\pi_{p..b}$ for the subcurve $(\pi_p,\pi_{p+1},\ldots,\pi_b)$. 
By interpolating between vertices, we can also view $\pi$ as a continuous function $\pi\colon [1,n] \to \mathbb{R}^2$ with $\pi_{i+\lambda} = (1-\lambda) \pi_i + \lambda \pi_{i+1}$ for $i \in \{1,\ldots,n-1\}$ and $\lambda \in [0,1]$.

Formally, the \Fr distance is defined as follows.
Let $\Phi_n$ be the set of all continuous and non-decreasing functions $\phi$ from $[0,1]$ onto $[1,n]$. Then two curves $\pi,\sigma$ have \Fr distance
$$ \dfr(\pi,\sigma) := \inf_{\substack{\phi_1 \in \Phi_{|\pi|} \\\phi_2 \in \Phi_{|\sigma|}}} \max_{t \in [0,1]} \|\pi_{\phi_1(t)} - \sigma_{\phi_2(t)}\|. $$
We call $\phi := (\phi_1,\phi_2)$ a \emph{traversal} of $(\pi,\sigma)$.

\section{A Quadtree Finds Candidates} \label{sec:quadtree}

Consider two curves $\pi,\sigma$. Since any traversal starts at both starting points, we have $\|\pi_1 - \sigma_1\| \le \dfr(\pi,\sigma)$. Similarly, since any traversal ends at both endpoints, we have $\|\pi_{|\pi|} - \sigma_{|\sigma|}\| \le \dfr(\pi,\sigma)$. 

Denote by $\textup{min-}x(\pi)$ the minimal $x$-coordinate of any vertex of curve $\pi$. Then $|\textup{min-}x(\pi) - \textup{min-}x(\sigma)| \le \dfr(\pi,\sigma)$. Indeed, 
if $\textup{min-}x(\sigma) \ge \textup{min-}x(\pi)$ then the vertex with minimal $x$-coordinate on $\pi$ has distance at least $\textup{min-}x(\sigma)-\textup{min-}x(\pi)$ to any vertex of $\sigma$, and the remaining case symmetrically yields the lower bound $\textup{min-}x(\pi)-\textup{min-}x(\sigma)$. We obtain three symmetric lower bounds by considering maximal $x$-coordinates ($\textup{max-}x(\pi)$), minimal $y$-coordinates ($\textup{min-}y(\pi)$), and maximal $y$-coordinates ($\textup{max-}y(\pi)$). In total, we obtain:
\begin{align*}
  \dfr(\pi,\sigma) \ge&\; LB_{\textup{F}}(\pi,\sigma), \text{ with} \\
   LB_{\textup{F}}(\pi,\sigma) :=&\, \max\big\{ \|\pi_1 - \sigma_1\|, \|\pi_{|\pi|} - \sigma_{|\sigma|}\|, \\
  & |\textup{min-}x(\pi) - \textup{min-}x(\sigma)|,  |\textup{max-}x(\pi) - \textup{max-}x(\sigma)|, \\
  & |\textup{min-}y(\pi) - \textup{min-}y(\sigma)|,  |\textup{max-}y(\pi) - \textup{max-}y(\sigma)| \big\}. \\
\end{align*}

In the first phase of our algorithm we determine the candidate set $D' := \{\sigma \in D \mid LB_{\textup{F}}(\pi,\sigma) \le \delta \}$. Note that for any remaining curve $\sigma \in D \setminus D'$ we have $\dfr(\pi,\sigma) \ge LB_{\textup{F}}(\pi,\sigma) > \delta$, so there are no false negatives.

In order to determine $D'$,
for each curve $\sigma \in D$ we store an 8-dimensional vector consisting of the coordinates for the start and endpoints as well as the largest and smallest coordinates in both dimensions. We store all these vectors in an 8-dimensional analogue of a quadtree.
Given a query $(\pi,\delta)$, for the query curve $\pi$ we also compute this 8-dimensional vector. 
Then using the quadtree we can enumerate all curves $\sigma \in D'$.

Because of feature (2) from the introduction, curves $\sigma \in D'$ typically also have $\dfr(\pi,\sigma)$ not much larger than $\delta$. This means that this enumeration has not too many false positives in practical situations, i.e., there are not too many curves $\sigma \in D'$ with $\dfr(\pi,\sigma) > \delta$, see also the last line of Table~\ref{tab:1}.

\medskip

On all enumerated curves $\sigma \in D'$, we first run the filters from the next section, and only if these heuristics fail we run the exact decision algorithm from Section~\ref{sec:exact}. This determines for each curve $\sigma \in D'$ whether $\dfr(\pi,\sigma) \le \delta$ and thus solves the range searching problem.

\section{Positive and Negative Filters} \label{sec:filters}

Let $\pi,\sigma$ be curves with $n,m$ vertices, respectively, and let $\delta > 0$. Our task is to decide whether $\dfr(\pi,\sigma) \le \delta$.
We make use of the following heuristics.

\paragraph{Greedy Algorithm}
We first run a simple greedy algorithm, which was previously discussed in~\cite{BringmannM16}: We start at $i=1, j=1$. When we are at $(i,j)$, we either increment $i$ or $j$ or both, specifically we go to the pair $(i',j') \in \{(i+1,j), (i,j+1), (i+1,j+1)\}$ minimizing the distance $\| \pi_{i'} - \sigma_{j'} \|$. From these three possible steps we ignore the ones that would increase $i$ beyond $n$ or $j$ beyond $m$. We end once we reach $(n,m)$. If during the whole process we stayed within distance $\delta$, then it follows that $\dfr(\pi,\sigma) \le \delta$. This easy and very efficient ($O(n+m)$ time) algorithm filters out some curves $\sigma \in D'$ with $\dfr(\pi,\sigma) \le \delta$.

\paragraph{Negative Filter}
Set $q_1 := 1$. For any $i \in \{2,\ldots,n\}$, let $q_i$ be the first point on $\sigma$ after $q_{i-1}$ that is within distance $\delta$ of $\pi_i$, i.e., set $q_i := \min\{q \in [q_{i-1},m] \mid \| \sigma_q - \pi_i \| \le \delta \}$. Note that here $q$ may be a real number, i.e., $\sigma_q$ may be an interpolated point. An easy inductive proof shows that in any traversal that stays in distance~$\delta$, any point $\sigma_{z}$ that is visited while being at $\pi_i$ satisfies $z \ge q_i$, i.e., $q_i$ is a lower bound on the position in $\sigma$ while being at $\pi_i$. Hence, if it happens that some value $q_i$ is undefined (since the set that we minimize over is empty), then we can conclude $\dfr(\pi,\sigma) > \delta$. This filters out some curves $\sigma \in D'$ that have too large \Fr distance.

The above procedure can be implemented in time $O(n+m)$. However, the constant factors are relatively large, since in each step we compute the intersection of a circle ($\{ x \in \mathbb{R}^2 \mid \| x - \pi_i \| = \delta \}$) and a line segment ($\{\sigma_q \mid q \in [j,j+1]\}$). Hence, we instead implemented a discrete version of this filter, where intuitively we aim to round down each $q_i$ to an integer, so that we only consider vertices in both $\pi$ and $\sigma$. Formally, we replace the definition of $q_i$ by $\tilde q_i := \min\{j \in \{\tilde q_{i-1},\ldots,m\} \mid \| \sigma_j - \pi_i \| - \| \sigma_j - \sigma_{j+1}\| \le \delta \}$. Observe that we relax the distance constraint by the length of the current segment in $\sigma$, which indeed implies $\tilde q_i \le q_i$. Thus, when it happens that $\tilde q_i$ is undefined, we can conclude that $\dfr(\pi,\sigma) > \delta$.

As this procedure is not symmetric in $\pi$ and $\sigma$, we can run it a second time with their roles swapped.

\section{Fr\'echet Distance Decider} \label{sec:exact}

It remains to handle the curves $\sigma \in D'$ reported by the first phase (quadtree) that are not filtered out in the second phase (filters). 
Preliminary experiments show that this third step dominates the running time compared to the first two phases, see Table~\ref{tab:1}, which is why we spent most of our development time on an optimized implementation of this phase.

We start by reviewing the standard free-space diagram algorithm for the \Fr distance. Then we describe our intuition for an optimized implementation, and finally we present the details of our new algorithm. In the whole section, let $\pi,\sigma$ be curves with $n,m$ vertices, respectively, and let $\delta > 0$. 

\subsection{Standard Free-space Algorithm}

The set $F := \{ (p,q) \in [1,n] \times [1,m] \mid \| \pi_p - \sigma_q\| \le \delta\}$, which describes the pairs $(p,q)$ such that $\pi_p$ and $\sigma_q$ are in distance $\delta$, is called the \emph{free-space diagram} of $\pi$ and $\sigma$. 
We also define the \emph{reachable free-space} $R \subseteq F$ as the pairs $(p,q)$ such that there exists an $x$- and $y$-monotone path through the free-space diagram from the lower left corner $(1,1)$ to $(p,q)$. 
These concepts are used in all known algorithms for the \Fr distance, because of the fact that $\dfr(\pi,\sigma) \le \delta$ holds if and only if $(n,m) \in R$.

A subset $C_{i,j} := F \cap [i,i+1] \times [j,j+1]$ for $i \in \{1,\ldots,n-1\}, j \in \{1,\ldots,m-1\}$ is called a \emph{free-space cell}. 
It is a well-known fact that each $C_{i,j}$ is the intersection of an ellipse with the box $[i,i+1] \times [j,j+1]$, in particular $C_{i,j}$ is convex. We denote by $F^l_{i,j}$ the left boundary of $C_{i,j}$, i.e., the interval $F \cap [i,i+1]\times \{j\}$, and similarly by $F^b_{i,j}$ the bottom interval $F \cap \{i\} \times [j,j+1]$. Similarly, $F^r_{i,j} := F^l_{i,j+1}$ and $F^t_{i,j} := F^b_{i+1,j}$. Any such  interval can be computed in constant time by one intersection of a circle and a line segment.

Furthermore, we define the reachable subset $R^l_{i,j} := F^l_{i,j} \cap R$ and similarly $R^b_{i,j}, R^r_{i,j}, R^t_{i,j}$. Then $\dfr(\pi,\sigma) \le \delta$ holds if and only if $(n,m) \in R^t_{n-1,m-1}$. 
A simple procedure computes the output intervals $R^r_{i,j}$ and $R^t_{i,j}$ from the input intervals $R^l_{i,j}$ and $R^b_{i,j}$ in constant time. Since also the leftmost boundaries $R^l_{i,1}$ and bottommost boundaries $R^b_{1,j}$ can each be computed in constant time, we can determine all reachability intervals in total time $O(nm)$. This is the standard algorithm for deciding whether $\dfr(\pi,\sigma) \le \delta$. 

\subsection{Intuition for our Algorithm}

For intuition, think of $\pi$ and $\sigma$ being equal curves, specifically a line segment of length $L$ subdivided $n$ times.
In this very simple case, the reachable free-space is a tube of the form $R = \{ (p,q) \in [1,n] \times [1,m] \mid |p-q| \le \Delta \}$ with $\Delta = \delta n / L$. Our intuition is that up to a ``smooth'' transformation and some ``noise'' at the boundary of $R$, this case models practical input curves quite well. 

Observe that in this specific case $R$ consists of $\Theta(\Delta n)$ non-empty free-space cells. An algorithm enumerating all non-empty free-space cells thus takes time $\Omega(\Delta n)$. However, there are large blocks of the reachable free-space that are either completely filled (i.e., reachable) or completely empty (i.e., unreachable). Let $B$ denote the number of ``boundary'' free-space cells, which are neither completely filled nor completely empty. In our exemplary situation we have $B = O(n)$.
By an analogy to (standard 2-dimensional) quadtrees, it can be shown that there is a partitioning of $[1,n] \times [1,m]$ into $O(B)$ blocks of the form $[p,b] \times [q,d]$ such that each block is either completely filled or completely empty or a boundary free-space cell. The goal of our recursive algorithm is to compute such a partitioning. 
In the motivating example of $\pi$ and $\sigma$ being the same subdivided straight line, this complexity is $O(B) = O(n)$, which yields a speedup compared to the time $\Omega(\Delta n)$ of the standard algorithm for large $\Delta$.

However, since checking whether a block is completely filled or completely empty is costly, we replace this step by a heuristic involving the triangle inequality. For this heuristic variant, we cannot prove a bound of $O(B)$ on size of the generated partitioning, but on practical data it behaves well.

\medskip

In a nutshell, the idea is to identify large blocks of completely filled or completely empty reachable free-space, and handle them efficiently (in constant time).

\subsection{Details of our Algorithm}

We construct a partitioning of the reachable free-space recursively as follows.
Given subcurves $\pi_{p..b}, \sigma_{q..d}$ and their input intervals $R^l_{i,q}$ for $i \in \{p,\ldots,b\}$ and $R^b_{p,j}$ for $j \in \{q,\ldots,d\}$, the recursive subproblem is to determine all output intervals $R^r_{i,d}$ for $i \in \{p,\ldots,b\}$ and $R^t_{b,j}$ for $j \in \{q,\ldots,d\}$. At the root of the recursion we have $[p,b] = [1,n]$ and $[q,d] = [1,m]$. 

We now describe how to handle a recursive subproblem.
Suppose that the lower left corner $(p,q)$ is in the input intervals.
Also suppose that the distances $\|\pi_i - \sigma_j\|$ are at most $\delta$ for all $i \in \{p,\ldots,b\}, j \in \{q,\ldots,d\}$. Then all free-space cells $C_{i,j}$ for $i \in \{p,\ldots,b-1\}, j \in \{q,\ldots,d-1\}$ are completely filled, and thus we are done with the subcurves $\pi_{p..b}, \sigma_{q..d}$. However, testing the condition requires iterating over all $i$'s and $j$'s and thus is too costly. Hence, we replace it with a cheaper condition: We test whether $\|\pi_p - \sigma_q\| + \|\pi_{p..b}\| + \|\sigma_{q..d}\| \le \delta$, i.e., whether the distance between the startpoints of the current subcurves plus the curvelengths of the current subcurves is at most $\delta$. Note that by triangle inequality this implies that all distances $\|\pi_i - \sigma_j\|$ are at most $\delta$ for any $i \in \{p,\ldots,b\}, j \in \{q,\ldots,d\}$, and thus the free-space spanned by $\pi_{p..b}$ and $\sigma_{q..d}$ is completely filled. 
Also, this check can be performed in constant time after the following preprocessing: For any curve in the database as well as for the query curve we store for each prefix $\pi_{1..i}$ the curvelength $\|\pi_{1..i}\|$. This allows us to determine the curvelength of $\pi_{p..b}$ as $\|\pi_{1..b}\| - \|\pi_{1..p}\|$ in constant time. This yields a quick test handling large completely filled blocks in constant time.

We use a similar test to determine whether the free-space spanned by $\pi_{p..b}$ and $\sigma_{q..d}$ is completely empty. 

In case these tests do not apply, we split one of the two curves into two halves, say we split $\pi_{p..b}$ into $\pi_{p..r}$ and $\pi_{r..b}$ for $r = \lfloor (p+b)/2 \rfloor$. We then recursively determine the reachable free-space in the subcurves $(\pi_{p..r}, \sigma_{q..d})$ and then in $(\pi_{r..b}, \sigma_{q..d})$. 

In case we reach a cell, i.e., $b = p+1$ and $d = q+1$, we handle it as in the usual free-space diagram algorithm. This finishes the description of our recursive algorithm. 

\medskip

The worst-case complexity of this algorithm is $O(n m \log (nm))$, however, this does not reflect the behaviour on realistic inputs.
We leave it as an open problem for the theory community to explain the efficiency of our algorithm on a reasonable model of curves.

\begin{sidewaystable*}[t]
  \centering
\begin{tabular}{ | l | c | c | c | c | c |}
  \hline			
  output size $k$: & 0 & 1 & 10 & 100 & 1000 \\
  \hline
  total time: & 0.44 $\pm$ 0.12 ms & 0.72 $\pm$ 0.47 ms & 2.00 $\pm$ 2.11 ms & 11.58 $\pm$ 15.97 ms & 60.44 $\pm$ 92.25 ms \\
  \hline
  \hline
  time for quadtree: & 0.42 $\pm$ 0.10 ms & 0.44 $\pm$ 0.10 ms & 0.45 $\pm$ 0.09 ms & 0.55 $\pm$ 0.12 ms & 0.77 $\pm$ 0.14 ms \\
  \hline
  time for greedy filter: & 0.01 $\pm$ 0.01 ms & 0.02  $\pm$ 0.02 ms & 0.07 $\pm$ 0.06 ms & 0.53 $\pm$ 0.32 ms & 5.39 $\pm$ 2.96 ms \\
  \hline
  time for negative filter: & 0.01 $\pm$ 0.02 ms & 0.04 $\pm$ 0.06 ms & 0.16 $\pm$ 0.20 ms & 0.92 $\pm$ 1.32 ms & 4.44 $\pm$ 7.33 ms \\
  \hline
  time for exact decider: & 0.01 $\pm$ 0.05 ms & 0.23 $\pm$ 0.40 ms & 1.31 $\pm$ 1.87 ms & 9.58 $\pm$ 14.44 ms & 49.77 $\pm$ 83.39 ms \\
  \hline
  \hline
  total without greedy filter: & 1.02 $\pm$ 0.48 ms & 1.52 $\pm$ 0.86 ms & 5.93 $\pm$ 4.25 ms & 46.55 $\pm$ 37.76 ms & 345.42 $\pm$ 290.81 ms \\
  \hline
  total without negative filter: & 0.44 $\pm$ 0.14 ms & 0.78 $\pm$ 0.66 ms & 2.56 $\pm$ 3.14 ms & 12.00 $\pm$ 15.71 ms & 65.74 $\pm$ 94.74 ms \\
  \hline
  total without exact decider: & 0.60 $\pm$ 1.42 ms & 2.58 $\pm$ 4.90 ms & 12.20 $\pm$ 21.72 ms & 91.16 $\pm$ 163.71 ms & 571.88 $\pm$ 1114.99 ms \\
  \hline  
  \hline  
  $\#$ false positives of quadtree: & 0.16 $\pm$ 0.53 & 1.06 $\pm$ 1.76 & 4.05 $\pm$ 5.92 & 9.63 $\pm$ 13.48 & 69.43 $\pm$ 171.58 \\
  \hline  
\end{tabular}
  \caption{Experimental evaluation on the GIS Cup example dataset.}
  \label{tab:1}
  \vspace{-0.3cm}
\end{sidewaystable*}

\section{Experimental Results} \label{sec:results}

A set of $\approx\! 20\,000$ trajectories was provided by the GIS Cup organizers as test data.
They each have between $11$ and $769$ vertices. 
We performed preliminary experiments on this database on a MacBook Air with an 1,8 GHz Intel Core i5 and 8GB RAM.
The time to read the database from the hard disc is about 9700ms, which dominates the time of 310ms to build our data structures. 

On this database $D$, we generate random queries as follows. Fix an outputsize $k \in \{0,1,10,100,1000\}$. We pick a random curve $\pi \in D$. Using a binary-search-like procedure, we determine a random distance threshold $\delta$ such that there are exactly $k$ curves in $D$ within distance $\delta$ to $\pi$. (For some curves $\pi \in D$ such a threshold does not exist since there are multiple curves with the same distance to $\pi$; in this case we ignore $\pi$ and pick a new random curve.) We then measure the running time for the query $(\pi,\delta)$. 

The results are shown in Table~\ref{tab:1}, where each entry gives the mean and standard deviation over 500 such random queries. 
The first line shows the time per query.
Fitting this first line to the model $a + k^b \cdot c$ yields an approximation of $\approx\! 0.44\textup{ms}+k^{0.79} \cdot 0.27$ms for the average query time in terms of the outputsize $k$. 

The next four lines show the time spent for our main components. Note that for $k \ge 10$ the dominant part is the exact decision procedure, while for smaller $k$ our solution could possibly benefit from an improved quadtree implementation. 

The next lines show the average query time if we (1) remove the greedy filter, (2) remove the negative filter, or (3) replace our novel decision procedure with the standard decision algorithm for the \Fr distance. Note that all components are necessary as the running time increases by at least a factor $\approx\! 2$ -- except for the negative filter, but we believe that the latter is benefitial for other types of data. 

Finally, the last line of Table~\ref{tab:1} shows the number $p$ of false positives of our quadtree data structure, i.e., while the correct output size is $k$ the quadtree returns $k+p$ candidates. Note that this number $p$ is very small on average compared to $k$, showing that our lower bound $LB_{\textup{F}}(\pi,\sigma)$ and the resulting quadtree data structure are very effective. (Recall that these false positives are catched by our filters or exact decider; the complete algorithm has no false positives or negatives.)

\section{Conclusion}

We described a novel implementation for near-neighbors queries with respect to the \Fr distance, which won the ACM SIGSPATIAL GIS Cup 2017. 
It remains to incorporate this or a similarly fast implementation into one of the standard libraries for geographic information systems, in order to make it usable in a wide range of applications.

We leave it as an open problem to theoretically explain the running time of our algorithm on a suitable model of realistic input curves.
The biggest issue for this is that it seems like our algorithm uses some features that are present in many realistic input curves, but which are not guaranteed by the standard models of realistic input curves used in computational geometry. One such property is that curves with similar start- and endpoints have small \Fr distance.

\bibliographystyle{plain}
\bibliography{frechet} 

\end{document}